\documentclass[prb,twocolumn,superscriptaddress,showpacs]{revtex4}
\usepackage{epsfig,color}

\begin{document}
\title{Exciton spectra
in vertical stacks of triple and quadruple quantum dots \\ in an
electric field}
\author{B. Szafran} \affiliation{Faculty of Physics and Applied
Computer Science, AGH University of Science and Technology, al.
Mickiewicza 30, 30-059 Krak\'ow, Poland}
\author{E. Barczyk} \affiliation{Faculty of Physics and Applied
Computer Science, AGH University of Science and Technology, al.
Mickiewicza 30, 30-059 Krak\'ow, Poland}
\author{F.M. Peeters}
\affiliation{Departement Fysica, Universiteit Antwerpen,
Groenenborgerlaan 171, B-2020 Antwerpen, Belgium}
\author{S. Bednarek} \affiliation{Faculty of Physics and Applied
Computer Science, AGH University of Science and Technology, al.
Mickiewicza 30, 30-059 Krak\'ow, Poland}

\date{\today}

\begin{abstract}
We study an electron-hole pair in a stack of multiple quantum dots
in the presence of an external electric field using the
configuration interaction approach. We find that the bright energy
levels can be grouped into families which are associated with the
hole localized in a specific dot of the stack. The exciton energy
levels undergo avoided crossings as function of the external
electric field with different pattern for each family. We show that
the variation of the depths of the dots along the stack can be
deduced from the exciton spectrum. In the strong confinement limit
the families are mixed by a weak electric field due to hole
tunneling. This results in a characteristic multiple avoided
crossing of energy levels belonging to different families with an
accompanying modulation of the recombination probabilities and an
appearance of a single particularly bright state. We discuss the
modification of the spectrum when dots are added to the stack.
\end{abstract}
 \maketitle

\section{Introduction}
Double quantum dots are considered promising for construction of
basic quantum information processing devices with electron
spins\cite{loss} or excitons\cite{molinari} confined in separate
dots acting as quantum bits interacting in a way that can be
controlled for instance by an external electric field. One technique
used for double dot formation uses the spontaneous tendency of the
self-assembled InGaAs quantum dots to stack one above the other
during epitaxial growth. \cite{Solomon,Ledentsov,stack1,stack2}
Coupling between the dots within a stacked was detected early in
photoluminescence experiments\cite{Solomon,Ledentsov,fafard} as red
shifts of the exciton energy lines that appear when the interdot
barrier thickness is reduced. Red shifts are also observed as
function of the number of dots forming the
stack.\cite{Solomon,Ledentsov} Only relatively recently measurements
of the photoluminescence spectrum with an external electric field
applied to a {\it pair} of dots were
performed.\cite{ek1,ek2,ek3,ek4,xk1,xk2,doty,small} These
measurements\cite{ek1,ek2,ek3,ek4,xk1,xk2,doty,small} exploit the
optical signatures of coupling between the dots to establish the
role of electron and hole tunneling as well as the interaction
between the carriers in the artificial molecules. They proved useful
for probing the confinement potential in asymmetric quantum
dots,\cite{ek3} examining the spin interactions in the fine
structure of the photoluminescence lines\cite{small} and the
observation of an electric-field tunable $g$ factor.\cite{doty}

In quantum information processing the idea of using a pair of
quantum dots as coupled quantum bits implies as a natural extension
an array of quantum dots as a scalable quantum register.\cite{qr}
Recently, there has been a growing interest in artificial molecules
formed by more than two dots. For instance, vertically coupled
multiple dots formed in a gated quantum wire\cite{samuelson} were
used for construction of a single-electron pump. It was demonstrated
\cite{klein} that a triple quantum dot can be used to construct
gates for multivalued logic.  Moreover, usage of three quantum dots
as a source of spin-entangled currents was demonstrated.\cite{se}
Stability diagram of an artificial molecule formed by three planar
dots created in a gated two-dimensional electron gas was
determined.\cite{hawrylal} Artificial molecules formed of multiple
self-assembled quantum dots in planar geometry were also
fabricated.\cite{mqd0,mqd1,mqd2}

Till now theoretical\cite{t3,t4,t5,t6,t7} and experimental work
\cite{ek1,ek2,ek3,ek4,xk1,xk2,doty,small} on the electric field
effect on photoluminescence spectra was limited to artificial
molecules formed by pairs of dots. The purpose of the present paper
is to extend the previous theoretical work on the exciton spectrum
of two vertically coupled quantum dots to a vertical stack of three
and four quantum dots. The studied objects are an intermediate stage
between the thoroughly studied case of single- and double dots and
the dot superlattice that is eventually formed with a large number
of stacked dots. We discuss the evolution of the spectra and
electron-hole correlations when subsequent dots are added to the
stack. According to the results presented below the spectra contain
spectacular anomalous multiple avoided crossings of bright energy
levels related to both the electron and hole tunneling which may be
readily experimentally verified since both sample growth formation
and measurement techniques are available.

The electric field breaks the electron tunnel coupling between the
dots and separates electrons from holes. For a pair of dots the
dissociation of the exciton by the removal of the electron from the
deeper dot leads to a characteristic avoided crossing of a bright
energy level and a dark one\cite{ek1,ek2} with spatially separated
carriers. Most of the experiments \cite{ek1,ek2,ek3,ek4,xk1,xk2}
were performed in the intermediate coupling regime with barrier
thickness that allows only the electron tunneling, with a hole
localized in a single dot. In such a case the optical signatures of
the interdot coupling are determined completely by only the electron
transfer \cite{ek1,ek2,ek3,ek4,xk1,xk2} induced by the electric
field. However, avoided crossings related to tunneling of the hole
are also observed \cite{doty} in the strong coupling limit for
barrier thickness of 2 nm. In the present paper we discuss both the
intermediate and strong coupling between the dots. We find that in
the intermediate coupling case the apparently complex spectra are
ordered by the hole which stays localized in a specific dot within
the stack. The position of the hole-containing dot in the stack
leads to an individual pattern of avoided crossings when the
extended electron states are manipulated by the electric field. This
should allow to probe the variation of the confinement potential
along the stack by means of photoluminescence measurements
 as previously applied for
intentionally grown asymmetric double quantum dots.\cite{ek3} In the
strong coupling limit we obtain avoided crossings that are due to
the hole tunneling, with a strong modulation of the recombination
probabilities favoring a single energy level in the center of the
crossing. Outside the avoided crossing range the hole becomes
localized in a single dot and the spectra evolve with the electric
field like in the intermediate coupling case.

Although the physics of the interdot coupling can be most
conveniently described by a model of identical dots, the dots that
are actually produced are never identical.\cite{Solomon,Ledentsov}
For a pair of dots one of them is always larger / deeper than the
other. For more than two dots a number of possibilities in the
variation of the depth of confinement potential along the stack is
possible. The purpose of the present paper is to extract the
features of the spectra that are independent of a specific order of
confinement potential depths along the stack.
 For that reason we consider
both systems of identical dots as well as dots with varied potential
depths. Of the latter we mainly exploit the realistic case in which
the depth of the confinement potential grows along the stack (below,
this case is referred to as a ''constant gradient``). This is
motivated by the fact that in the Stranski-Krastanov growth mode for
the same nominal number of InAs monolayers deposited for each layer
of dots, the size of the quantum dots tends to increase along the
stack.\cite{Solomon,Ledentsov}. However, since the potential of
separate dots can be intentionally modified\cite{ek3} during the
growth also other configurations will be considered in this paper,
namely those in which the dot with the strongest confinement is
situated inside the stack and not on its end.

We apply our configuration interaction approach introduced in
Ref.\cite{t3} and developed further in Ref.\cite{t6} which
successfully predicted\cite{t3} the mechanism of the dissociation of
the exciton \cite{ek1,ek2} and the trion \cite{ek4} by the electric
field.

\begin{figure}[ht!]
\hbox{\epsfysize=52mm
                \epsfbox[27 220 570 628] {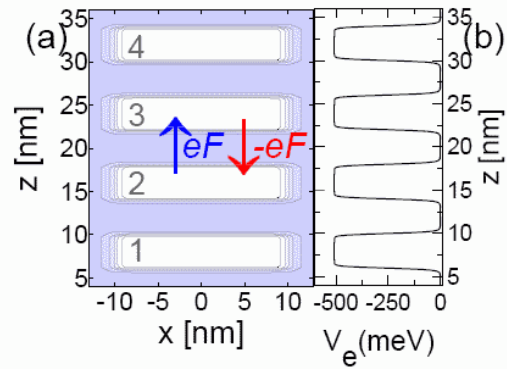}\hfill}
\caption{(color online) (a) The confinement potential (cross section
at $y=0$) for a stack of dots with interdot barrier of 4 nm -- the
lighter the shade of grey the deeper the potential. Dots have
identical size with diameter of 20 nm and height of 4 nm. The dots
are numbered from the lowest to the uppermost. The arrow at left
(right) shows the electric field force acting on the electron (hole)
for $F>0$. (b) The confinement potential for the electron plotted
along the axis of the stack ($x=0,y=0$).} \label{schema}
\end{figure}

The paper is organized as follows. In Section II we describe the
model. Section III contains the results for an ideal system of three
identical dots which are then generalized to the case of three and
four nonidentical dots. Results are discussed in Section IV.
Conclusions and summary are given in Section V.

\begin{figure}[ht!]
\hbox{\epsfysize=52mm
                \epsfbox[38 160 550 662] {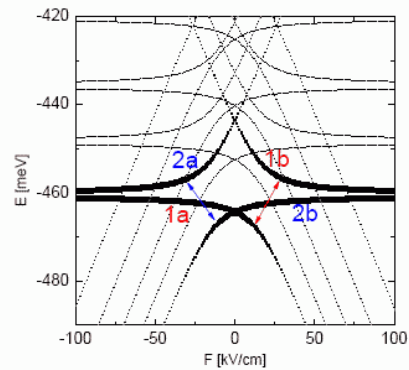}\hfill}
\caption{(color online) Exciton energy spectrum for two identical
dots separated by barrier thickness of 5 nm.  Thickness of the lines
is proportional to the recombination probability [See.
Eq.(\ref{recprob})]. The bright energy levels are denoted by the
number 1 or 2 which indicates the dot in which the hole is localized
(see Fig. \ref{schema}). The two arrows indicate the energy levels
that avoided cross when the electron is removed of the dot in which
the hole is localized. Letters ($a$,$b$) mark the bright energy
levels of the same family (with the hole localized in the same dot)
that appear for $F<<0$, $F\simeq 0$ and $F>>0$, respectively. }
\label{2e}
\end{figure}

\begin{figure}[ht!]
\hbox{\epsfysize=124mm
                \epsfbox[58 17 513 815] {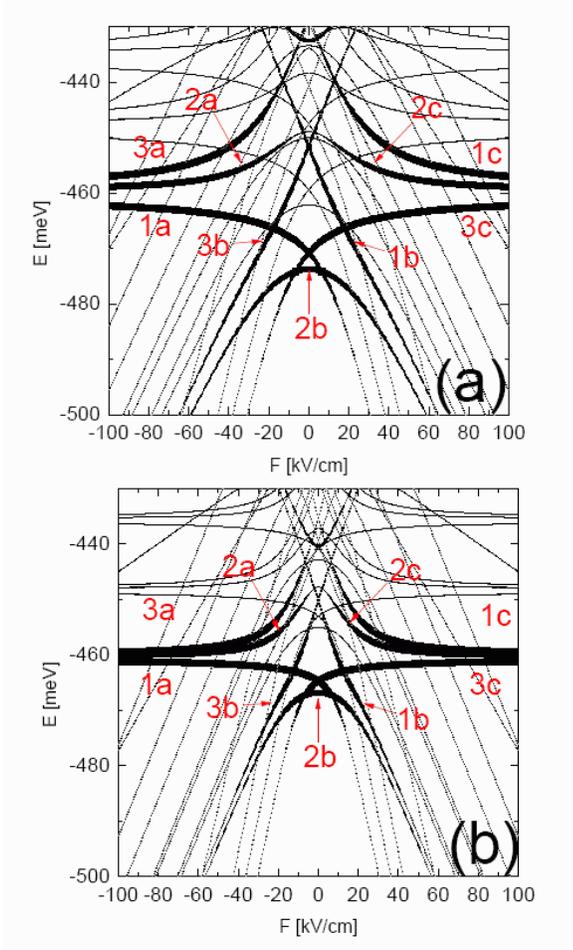}\hfill}
\caption{(color online) Exciton energy spectrum for three identical
dots separated by barrier thickness of 4 nm (a) and 5 nm (b).
Thickness of the lines is proportional to the recombination
probability [Eq. (\ref{recprob})]. The bright energy levels are
denoted by the  numbers 1, 2 or 3 which indicates the dot in which
the hole is localized (see Fig. \ref{schema}).  Letters
($a$,$b$,$c$) mark the bright energy levels of the same family (with
the hole localized in the same dot) that appear at $F<<0$, $F\simeq
0$ and $F>>0$, respectively. } \label{000_8}
\end{figure}

\section{Model and method}

\begin{figure}[ht!]
\hbox{\epsfxsize=60mm
                \epsfbox[94 24 515 816] {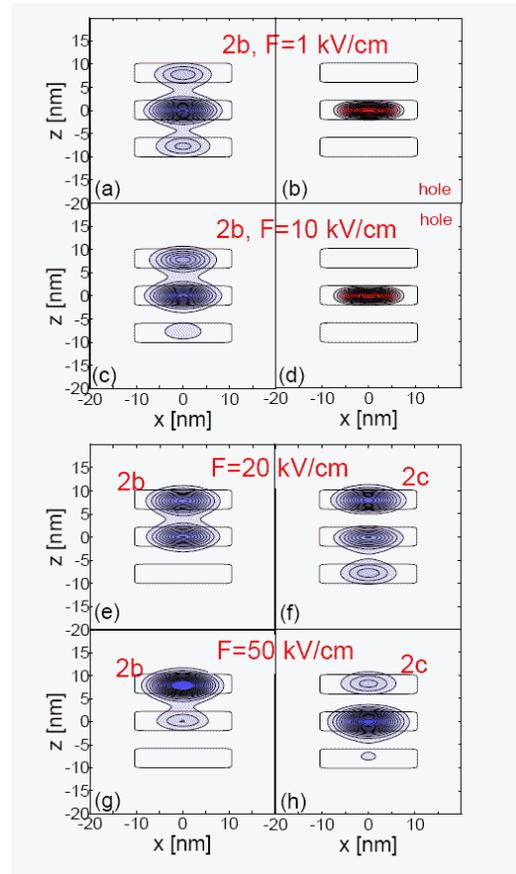}
                \hfill}
\caption{(color online) Cross section (y=0) of electron (a,c,e-h)
and hole (b,d) densities in the states of the $2x$ family, in which
the hole is localized in the central (2) dot for three identical
dots with interdot barrier thickness of 4 nm for various values of
the electric field $F$. The corresponding energy levels are plotted
in Fig. \ref{000_8}(a). Thin solid boxes indicate the region in
which the confinement potential is at least 10\% of its maximal
value with respect to the center of each dot. }\label{ff000_8}
\end{figure}

\begin{figure}[ht!]
\hbox{\epsfxsize=60mm
                \epsfbox[151 272 439 562] {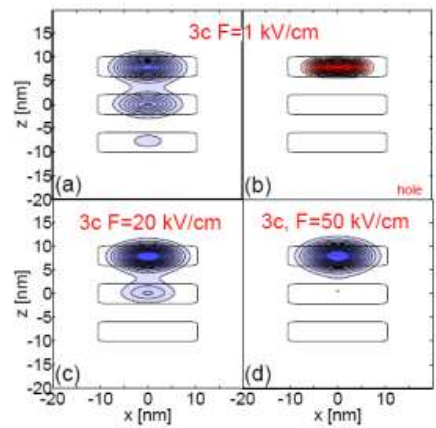}
                \hfill}
\caption{(color online) Same as Fig. \ref{ff000_8} but for the $3c$
energy level of Fig. \ref{000_8}. Plots (a,c,d) show the electron
density and (b) the hole density.}\label{ff000_83a}
\end{figure}

The model and the computational method are those as were presented
in Ref. \cite{t6} for two coupled dots. This approach accounts for
the electron-hole correlation in the external electric field
assuming the single valence band approximation and a simple
confinement potential model. The single band approximation for the
hole is justified by a small height of the dot which excludes the
light hole contribution from the lowest energy states. The dipole
moment induced by the charge transfer between the dots is much
larger than the one resulting from the wave function deformation
within a single dot. Therefore, we choose to apply a simple
confinement potential model. For a single dot we use a disk quantum
well,
\begin{equation}
V({\bf r})=-V_0/ \left[\left(1+\frac{x^2+y^2}{R^2}\right)^{10}
\left(1+\frac{(z-\zeta)^2}{Z^2}\right)^{10}\right], \label{pot}
\end{equation}
where $\zeta$ is the position of the center  of the dot along the
$z$ axis. The profile of the confinement potential is plotted by
shades of grey in Fig. \ref{schema}(a) for a stack of four identical
dots and Fig. \ref{schema}(b) presents the electron confinement
potential along the axis of the stack.

The dots are assumed to be perfectly aligned [see Fig.
\ref{schema}(a)]. A non-perfect alignment has a negligible effect on
the spectra when the electric field is oriented in the growth
direction.\cite{t6} Following the experimental data of Ref.
\cite{ek4} we take the diameter of the dots $2R$ equal to 20 nm and
the height $2Z$ to 4 nm. The effective masses ($m_e=0.037m_0$ for
the electron and  $m_h=0.45m_0$ for the hole) as well as the
dielectric constant ($\epsilon=12.5$) are taken for
In$_{0.66}$Ga$_{0.34}$As quantum dot \cite{t6} embedded in GaAs and
consequently the depth of the wells for the electron and hole are
respectively $V_0^e=508$ meV and $V_0^h=218$ meV. A variation of the
confinement potential within the stack is modeled by introducing
variations in the depths of the dots. We consider the following
Hamiltonian for an electron-hole pair
\begin{eqnarray}
H&=&-\frac{\hbar^2}{2m_e}\nabla_e^2-\frac{\hbar^2}{2m_h}\nabla_h^2+V_e({\bf
r}_e)+V_h({\bf r}_h)\nonumber \\ &&-\frac{e^2}{4\pi \epsilon
\epsilon_0 |{\bf r}_e-{\bf r}_h|}-e{\bf F} \cdot \left({\bf r}_e
-{\bf r}_h\right),
\end{eqnarray}
where ${\bf r}_e$, ${\bf r}_h$ are the electron and the hole
coordinates respectively, the confinement potentials for the
electron and hole are given by a sum of single-dot potentials of the
form given by Eq. (\ref{pot}),
 and ${\bf F}$ is the electric field. In this
Hamiltonian the energy is calculated with respect to the bottom of
the conduction  and the top of the valence band of the barrier
material (GaAs). Spin effects for the exciton, the electron-hole
exchange and the fine structure splitting are small\cite{small} at
the energy scale of the avoided crossings and are therefore
neglected in this paper.

The electron-hole pair eigenproblem is solved using the
configuration-interaction method with the single-particle
eigenfunctions ($f^{(j)}_{p}({\bf r}_p)$ where $j$ enumerates the
single-particle eigenfunctions and the particle notation is $p=e$
for the electron and $p=h$ for the hole) diagonalized in the
multicenter basis of Gaussian wave functions\cite{t6}
\begin{eqnarray}
f^{(j)}_{p}({\bf r}_p)&=& \sum_{i} c_i^{(j,p)} \exp\left[-\alpha_i^p
\left((x_p-x_i^p)^2+(y_p-y_i^p)^2\right)\nonumber \right. \\
&&\left. -\beta_i^p (z_p-z_i^p)^2\right],
\end{eqnarray}
where $c_i^{(j,p)}$ are the linear variational parameters and
$\alpha_i^p$, $\beta_i^p$ are the nonlinear variational parameters
describing the localization strength of $i^{\mathrm th}$ Gaussian
around point $(x_i^p,y_i^p,z_i^p)$.  For each dot we apply 11
Gaussian functions: 8 Gaussians on a circle around the axis of the
stack and 3 Gaussians along the axis. Exact positions of the
Gaussians and strength of their localization in the growth direction
($\beta_i^p$) and in the perpendicular plane ($\alpha_i^p$) are
optimized variationally, separately for the electron and for the
hole. The basis accounts for confined single-particle states with
angular momentum up to $\pm4\hbar$. Inclusion of more Gaussian
centers in the basis does not significantly improve the results. In
total, for a stack of $N$ dots the basis for the electron-hole pair
contains $(11N)^2$ localized wave functions given by products of
single-electron $f_e^{(j)}$ and single-hole functions $f_h^{(k)}$
\begin{equation}
\phi({\bf r}_e,{\bf r}_h)=\sum_{j=1}^{11N}\sum_{k=1}^{11N} d_{jk}
f_e^{(j)}({\bf r}_e)f_h^{(k)}({\bf r}_h).
\end{equation}
We discuss the electron $n_e({\bf r}_e)$ and hole $n_h({\bf r}_h)$
densities which are extracted from the exciton wave function $\phi$
whose arguments are vectors in six-dimensional space by integrating
its square over the coordinates of the other particle, i.e.
\begin{equation} n_e({\bf r}_e)=\int d{\bf r}_h
|\phi({\bf r}_e,{\bf r}_h)|^2
\end{equation}
and
\begin{equation} n_h({\bf r}_h)=\int d{\bf r}_e
|\phi({\bf r}_e,{\bf r}_h)|^2.
\end{equation}
Finally, the recombination probability for the exciton state wave
function $\phi$ is calculated as
\begin{equation} p=\left |\int d^3{\bf r}_e\int d^3{\bf r}_h \phi({\bf
r}_e,{\bf r}_h)\delta^3({\bf r}_e-{\bf r}_h) \right |^2 .
\label{recprob}
\end{equation}

\begin{figure*}[ht!]
\hbox{\epsfxsize=110mm
                \epsfbox[68 227 537 615] {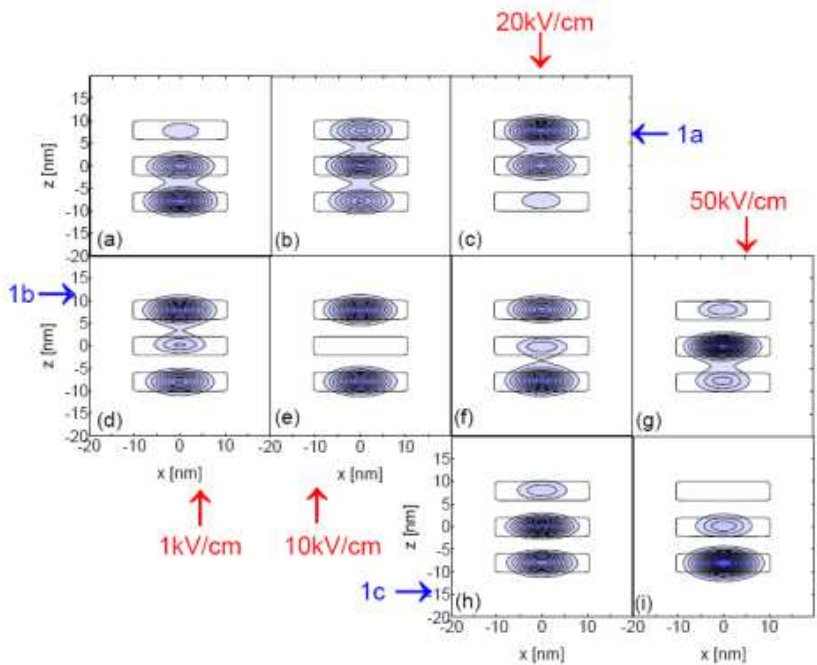}
                \hfill}
\caption{(color online) Same as Fig. \ref{ff000_8} but for the $1x$
family of states in which the hole is localized in the lowest dot.
All the plots show the electron density. The upper (a-c), middle
(d-g), and lower (h-i) row of plots correspond to $1a$, $1b$ and
$1c$ energy levels of Fig. \ref{000_8}. The columns of plots from
left to right correspond to $F=1$, 10, 20 and 50
kV/cm.\label{ff000_81x}}
\end{figure*}

At the end of this Section we would like to point out that the
present model is applied to a single stack of tunnel-coupled dots.
Quantum dots grown in arrays may interact with one another through
the electromagnetic field even though there is no tunneling between
the dots. Electromagnetic coupling effects include stimulated
emission amplification,\cite{ld1,ld2} polarization splitting of the
gain band\cite{rusin} observed at high excitation intensity in the
arrays of vertically stacked dots, and superradiance
observed\cite{superradiance} for spontaneous emission at weak
excitations. Experimentally, separation of a single stack of dots
can be performed for instance by etching a small mesa of the entire
sample.\cite{superradiance}

\section{Results}

In the present paper we discuss a stack of intermediately and
strongly coupled dots (see Section I). The strength of the coupling
can be conveniently quantified by the energy splitting of binding
and antibinding states for a pair of identical dots. For
intermediate coupling the electron splitting is of order 10 meV. The
hole tunneling remains negligible, with the energy splitting smaller
than 100 $\mu$eV. The results presented below for the intermediate
coupling regime were obtained for barrier thickness $t$ between 4
and 6 nm. For a {\it pair} of identical dots we obtain the electron
splitting of 9.5 meV and 27.5 meV for $t=6$ and 4 nm, respectively.
Corresponding hole energy splitting at these values of $t$ is 4
$\mu$eV and 70 $\mu$eV. As a strong coupling case we discuss spectra
obtained for $t=1$ nm thick barrier. For $t=1$ nm the electron
splitting is as large as 136 meV, which largely exceeds a possible
unintentional variation of the confinement depths between the dots.
This is not the case for the hole, for which the tunnel coupling
becomes significant but the energy splitting equals only 5.8 meV.
For the applied potential parameters the ground-state energy of the
exciton in a single quantum dot equals -460.7 meV. The ground state
becomes red-shifted with the addition of subsequent {\it identical}
dots to the stack: for interdot barrier of 1 nm we obtain
ground-state energies of -524.4, -546.8 and -556.2 meV for the stack
of two-, three-, and four-dots, respectively. The corresponding
relative shifts with addition of second, third and fourth dots are
equal to 63.8, 22.4, and 9.4 meV. With the addition of subsequent
dots the spectrum tends towards the superlattice limit. This is the
reason why the red shift reduces with addition of dots to the stack.
The red shift of the ground-state energy as well as the obtained
reduction of the relative shifts with the number of dots is in
qualitative agreement with the experimental results given for a
sample containing strongly-coupled stacks of several dots (see Fig.
4 of Ref. \cite{Ledentsov}). However, in the experimental work
\cite{Ledentsov} the corresponding relative shifts are 64.1, 34.3,
and 32.8 meV, i.e. they reduce with the number of dots in a less
pronounced manner than in our results for identical dots. This may
be for instance due to the fact that in this
experiment\cite{Ledentsov} the size of the dots grows as they are
added on top of the stack.

For our discussion of the evolution of the photoluminescence
spectrum with increasing number of dots within the stack it is
useful to inspect the spectrum for a couple of identical dots. This
is given in  Fig. \ref{2e} for intermediate coupling ($t=5$ nm) as
function of the electric field $F$. The spectrum contains a set of
bright energy levels [the thickness of the curves is set
proportional to the recombination probability calculated according
to Eq. (\ref{recprob})] and a multitude of dark energy levels
corresponding to hole excitations with non-zero angular momenta and
/ or separated carriers. For intermediate coupling the bright levels
can be divided into families which differ by the dot in which the
hole is localized. The families are labeled by numbers that indicate
the hole-containing dot in accordance with the order indicated
schematically in  Fig. \ref{schema}(a). Within each family we
additionally label the energy levels that appear as bright when the
electric field is swept from strong negative values to strong
positive values by subsequent letters $a$, $b$, $c$, etc.

Energy levels of the same family are involved in the avoided
crossings when the electric field removes the electron from the dot
occupied by the hole. For the dot couple the energy level $1a$ is
avoided crossed by energy level $1b$ (see the avoided crossings
marked by arrows in Fig. \ref{2e}) when the electron is removed of
the lower dot by a positive electric field $F>0$. A mirror
reflection of this avoided crossings appears between the energy
levels $2a$ and $2b$ at negative field. For intermediate coupling,
i.e. for negligible hole tunneling, the energy levels of different
families cross (see the crossing of $1a$ and $2b$ energy levels, as
well as the crossing of $2a$, $1b$ levels, both occurring at $F=0$
in Fig. \ref{2e}). In the ground state at $F=0$ such a crossing
leads to the ground-state degeneracy which is due to a negligible
hole tunneling. The problem of the ground state degeneracy for three
and four identical dots is discussed below.

\subsection{Three identical dots -- intermediate coupling}

Fig. \ref{000_8}(a) shows the energy spectrum for three identical
dots separated by tunnel barriers of 4 nm thick. The spectrum  is
perfectly symmetric with respect to the electric field orientation
and we will discuss only $F>0$. For $F=0$ the ground state ($2b$) is
nondegenerate -- in contrast to the two-dot case of Fig. \ref{2e} --
and corresponds to the hole localized in the central dot [see the
hole density plotted in Fig. \ref{ff000_8}(b)].\cite{odnosnik} The
electron occupies predominantly the same dot, but tunneling to the
neighbor dots is visible [see Fig. \ref{ff000_8}(a)]. For a
non-interacting electron-hole pair and negligible hole tunneling the
ground-state is {\it threefold} degenerate. Preference of the
electron to stay inside the central dot and the electron-hole
interaction leads to an effectively increased depth of the central
dot for the hole which lifts the ground-state degeneracy.

For $F$ increased from 0 to 10 kV/cm the electron density
corresponding to the $2b$ level is removed from the lowest dot and
increased in the uppermost dot [Fig. \ref{ff000_8}(c)].  The $2b$
level becomes dark near $F=50$ kV/cm [Fig. \ref{000_8}(a)] which is
due to the separation of the carriers [see Figs.
\ref{ff000_8}(e,g)]. The
 dissociation of the exciton in the $2b$ energy level at $F>0$ occurs
 via an avoided crossing with the $2c$ energy level.
 The electron in the $2c$ energy level [see Fig. \ref{ff000_8}(f)
and (h)] enters the central dot occupied by the hole when electron
density of $2b$  gets localized by the field in the uppermost dot
[see Figs. \ref{ff000_8}(f,h)].

Fig. \ref{ff000_83a}(a) shows that in the $3c$ energy level for
small field a residual presence of the electron in the lowest dot
(1) is observed. As the field increases the electron becomes
completely localized in the 3$^\mathrm{rd}$ dot (where the hole is
localized). This results in an increased intensity of the $3c$ line
with respect to $F=0$ that can be noticed in Fig. \ref{000_8}(a).
The $3x$ family has a more interesting behavior for $F<0$, and for a
symmetric dot this is a mirror reflection of the phenomena observed
for the $1x$ family at $F>0$. The energy levels $1a$ and $3c$ are
degenerate at $F=0$. At low field in the $1a$ state the electron
density is a reflection of the $3c$ density with respect to the
central dot [see Fig. \ref{ff000_81x}(a) and compare it to
\ref{ff000_83a}(a)]. In the $1a$ energy level at $F=0$ the
electron-hole distribution  has a non-zero electric dipole moment in
contrast to the $2b$ energy level. As a consequence, the $1a$ energy
level reacts to the field more strongly than $2b$ [see Fig.
\ref{000_8}(a)].  $1a$ first becomes the ground state energy level
and then it becomes dark [see Fig. \ref{000_8}(a)] when the electron
is removed from the lowest dot [see Figs. \ref{ff000_81x}(b,c)]. At
small electric field the electron in the $1b$ energy level is
localized mostly at the extreme dots of the triple stack [see Figs.
\ref{ff000_81x}(d-f)]. At higher field the electron in the $1b$
energy level becomes localized in the middle dot (the hole remains
in the lowest dot in all the states of the $1x$ family) and
consequently the energy level becomes dark. Note that the slope of
the $1b$ energy level when it becomes dark is twice smaller than for
the $1a$ ground state when it is dark. This is due to the electric
dipole moment which is twice smaller in $1b$ (hole in the lowest
dot, electron in the middle) than in the ground state $1a$ (hole in
the lowest dot, the electron in the uppermost). At high field the
electron becomes localized in the lowest dot only in the third state
($1c$) of the $1x$ family [see Figs. \ref{ff000_81x}(h,i)].

A strong electric field removes the electron tunnel coupling between
the dots, so the electron becomes localized in a single dot
similarly as the hole. In all the strong-field bright states $1c$,
$2c$, $3c$ the decrease of the electron energy with the electric
field is compensated by the increase of the hole energy. As a
consequence, at large $F$ all the bright levels $1c$, $2c$ and $3c$
become degenerate.  For increased barrier thickness to 5 nm
degeneracy appears for smaller $F$ [see Fig. \ref{000_8}(b)].

\begin{figure*}[ht!]
\hbox{\epsfxsize=160mm
                \epsfbox[19 328 577 508 ] {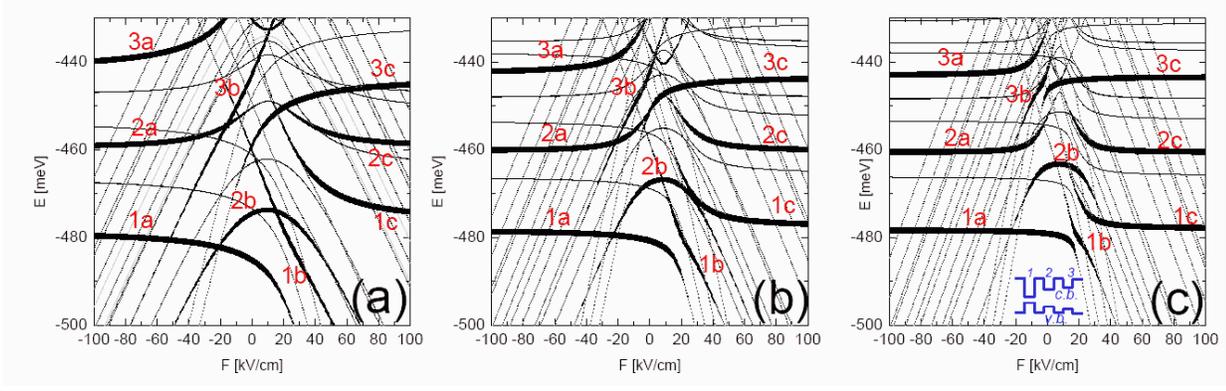}}
\caption{(color online) Same as Fig. \ref{000_8} but for three
nonidentical dots separated by barrier thickness of 4 nm (a) and 5
nm (b) and 6nm (c). The depth of the confinement potential of the
dots 1, 2, 3 is varied by +10 meV, 0, -10 meV (constant 'gradient'
of potential depth) respectively for both the electron and the hole.
Schematic drawing of the valence and conduction band along the stack
is presented in the inset of (c). } \label{non_10}
\end{figure*}

\begin{figure}[ht!]
\hbox{\epsfysize=52mm
                \epsfbox[35 174 555 662] {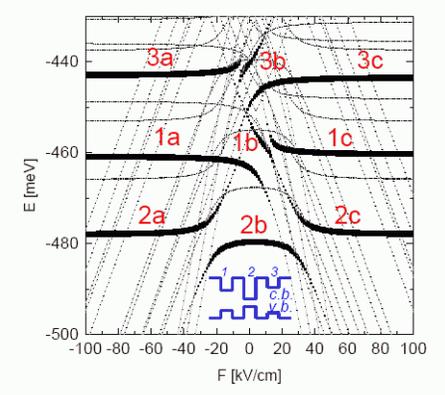}
                 \hfill}
\caption{(color online) Same as Fig. \ref{non_10}(c) but for varied
depths of the dot with barrier thickness of $t=6$ nm. The depth of
the confinement potential of the dots 1, 2, 3 is varied by 0, +10
meV, -10 meV, respectively for both the electron and the hole. The
inset shows a sketch of conduction and valence band extrema along
the growth direction. }\label{0mp}
\end{figure}

\begin{figure}[ht!]
\hbox{\epsfysize=52mm
                 \epsfbox[35 174 555 662]  {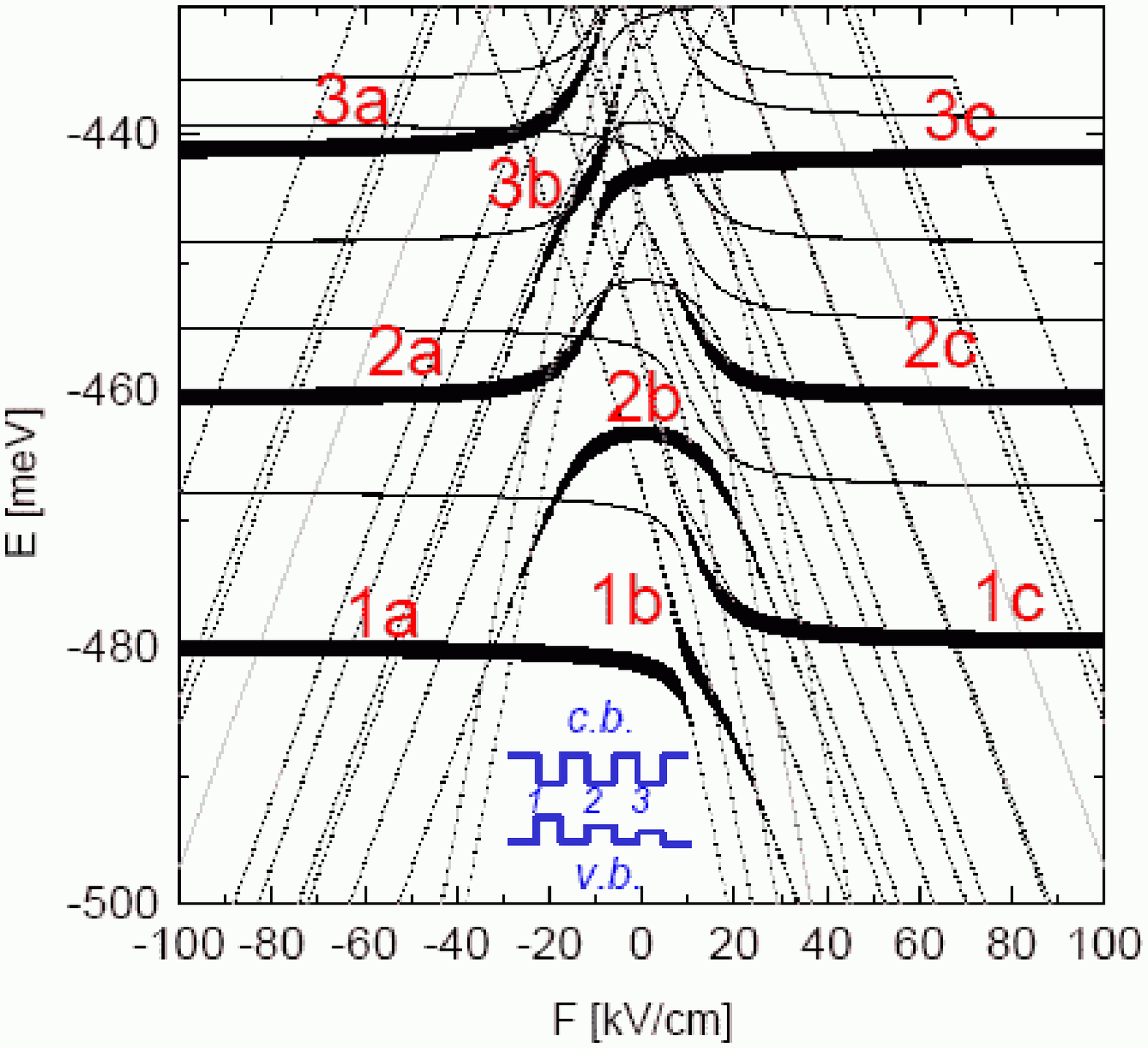}
                 \hfill}
\caption{(color online) Same as Fig. \ref{non_10}(c) but for varied
depths of the dot with barrier thickness of $t=6$ nm. The depth of
the confinement potential of the dots 1, 2, 3 is varied by +20 meV,
0, -20 meV, respectively for the hole. Dots have equal depth for the
electron. The inset shows a sketch of conduction and valence band
extrema along the growth direction. }\label{0h20}
\end{figure}

\begin{figure*}[ht!]
\hbox{\epsfysize=120mm
                \epsfbox[43 183 573 682] {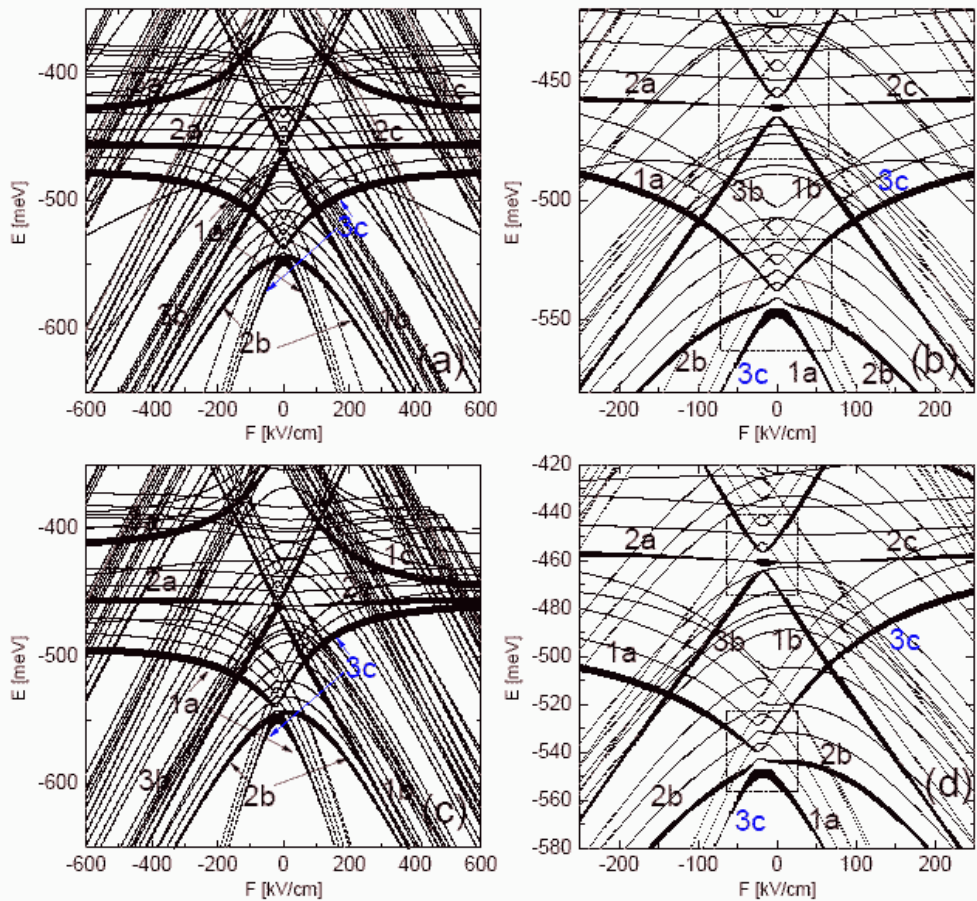}\hfill}

\caption{(color online) Spectra for the strong coupling case with
$t=1$ nm (a) correspond to identical dots and in (c) the depth of
the confinement potentials of the dots 1, 2, 3 is modified by +10, 0
meV, and -10 meV, respectively. (b) and (d) present fragments of (a)
and (c) respectively in which avoided crossings related to the
avoided crossings between different families are observed.}
\label{3strong}
\end{figure*}

\begin{figure*}[ht!]
\hbox{\epsfysize=162mm
                \epsfbox[24 84 567 816] {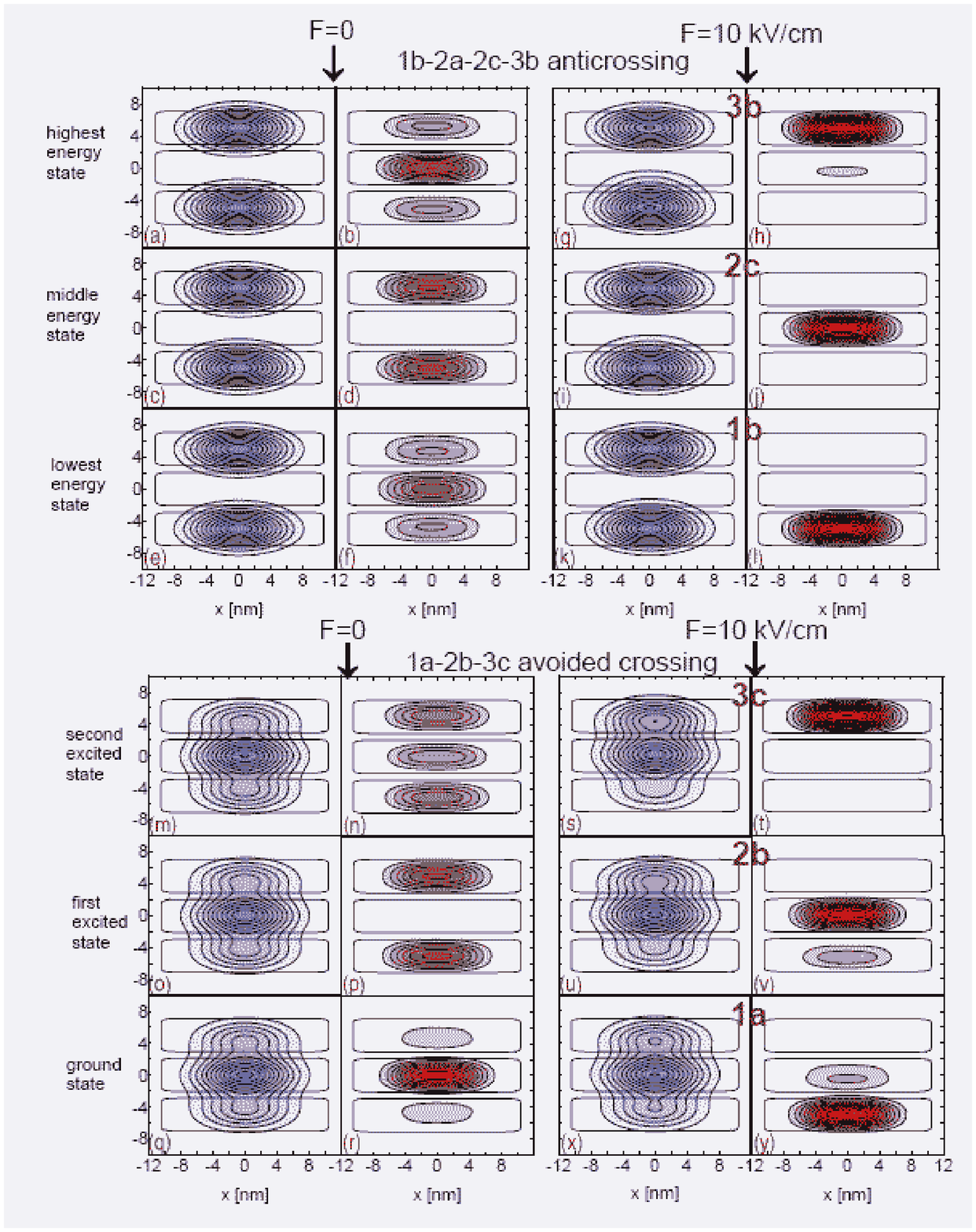}\hfill}
\caption{(color online) Cross section ($y=0$) of the electron and
hole densities for the energy levels participating in the two
avoided crossings presented in Fig. \ref{3strong}(b). The upper
group of plots (a-l) corresponds to the avoided crossing marked by
the upper rectangle in Fig. \ref{3strong}(b) and  the lower group
(m-y) to the ground-state avoided crossing marked by the lower
rectangle in Fig. \ref{3strong}(b). In each group - higher energy
states correspond to the upper panels. The first (second) column of
plots shows the electron (hole) density at $F=0$ (namely,
a,c,e,m,o,q for electron, and b,d,f,n,p,r for the hole). The third
(fourth) the electron (hole) density at $F=10$ kV/cm (g,i,k,s,u,x
for electron, and h,j,l,t,v,y, for the hole). Identities of the
states when their energy levels emerge from the avoided crossing can
be deduced from the plots for $F=10$ kV/cm. The densities are
therefore labeled according to the ``family'' notation used in
previous discussion.} \label{ff3strong}
\end{figure*}

The change of the spectrum with the addition of the third dot to the
two-dot stack can now be summarized by comparing of Fig. \ref{2e}
and Fig. \ref{000_8}. In the family $1x$ -- associated with the hole
localized in the lowest dot of the stack -- the bright energy levels
appearing at strong negative $F<<0$ and strong positive $F>>0$
fields -- $1a$ and $1c$, respectively -- exhibit an avoided crossing
(Fig. \ref{000_8}) like the $1a$ and the $1b$ energy levels for the
double dot case (Fig. \ref{2e}). Similarly as in the two-dot case
this avoided crossing -- related to the transfer of the electron
from the lower to the upper dot -- appears for $F>0$. The new
feature of the $1x$ family for three dots is the appearance of an
energy level ($1b$) which enters the $1a-1c$ avoided crossing nearly
linearly and appears as bright in the range of electric fields which
induces the avoided crossing. Family $3x$ associated with the dot at
the opposite extreme end of the stack is a mirror reflection of
$1x$. In comparison to the two-dot case a qualitatively new group of
bright levels is the $2x$ family, associated with the dot situated
{\it inside} the stack. The energy level which appears as bright
near $F=0$ ($2b$) has a parabolic dependence on $F$ and anticross
the two other bright energy levels of the family: $1a$ at $F<0$ and
$1c$ at $F>0$.

\subsection{Three nonidentical dots -- intermediate coupling}

Let us first consider the case when a constant ``gradient'' in the
depth of the confinement potential is present within the stack [see
the inset to Fig. \ref{non_10}(c)]. Fig. \ref{non_10} presents the
results for the case that the depth of the lowest dot (number 1) is
increased by 10 meV for both the electron and the hole while for the
uppermost dot (number 3) it is decreased by the same amount for both
carriers. The modification of the depth is directly translated into
relative energy shifts between the families of energy levels. For
large $F>0$, the $1c$, $2c$, and $3c$ energy levels are no longer
degenerate but they are separated by roughly 20 meV (sum of the
confinement energy difference for both carriers in the separate
dots).

The avoided crossings in the separate families are most clearly
visible for weaker coupling with $t=6$ nm in Fig. \ref{non_10}(c).
Although the families of energy levels are shifted, the qualitative
character of the avoided crossings within each family is the same as
in the stack of identical dots. In the $2x$ family a single avoided
crossing is observed at both electric field orientations and appears
when the electron is moved from the central to the upper or lower
dot. On the other hand for the $1x$ family only the positive
electric field induces avoided crossings. The first avoided crossing
corresponds to the removal of the electron from the lowest to the
uppermost dot and the second from the lowest to the middle dot.

The spectrum in case the central dot is the deepest is plotted in
Fig. \ref{0mp} for 6 nm thick tunnel barriers [see the inset to Fig.
\ref{0mp}, the depth of the central dot is increased by 10 meV, and
the uppermost is decreased by 10 meV]. Similarly as in the
``constant gradient'' case of Fig. \ref{non_10}(c) we see well
resolved patterns of the three families of energy levels known from
the results presented above. The families are shifted as expected
from the varied potential depths.

In the discussed calculations for nonidentical dots we assumed that
the depth of the dot is varied in the same way for both the electron
and the hole.  In Fig. \ref{0h20} we show the spectrum calculated
for equal electron confinement depths but with varied depths for the
hole: +20, 0, -20 meV along the growth direction. The adopted shifts
are equal to the sum of the electron and hole variation of Fig.
\ref{non_10}(c). We see that both spectra presented in Fig.
\ref{non_10}(c) and Fig. \ref{0h20} are nearly identical. Same
result is obtained for equal depth for the hole but varied potential
depth for the electron. In our previous work \cite{t3} on the
coupled two dot system we demonstrated that the electron-hole
interaction translates the asymmetry for one type of particle to an
effective confinement potential for the other. This conclusion still
holds for triple stacked dots.

\subsection{Three stacked dots -- strong coupling}

Fig. \ref{3strong}(a) shows the spectra for a stack of triple
identical dots separated by a $1$ nm thin barrier.  The strong
coupling between the dots not only increases the electron energy
splitting (which widens the spectrum in the energy scale) but also
makes the electron states less susceptible to the manipulation by
the external field. Therefore, in order to obtain an evolution of
the spectrum comparable to the one observed in the intermediate
coupling case for $t=4$ nm [Fig. \ref{000_8}(a)] both the electric
field and the energy scale had to be increased by a factor of about
5. With this scaling the pattern of energy levels is similar to the
intermediate-coupling spectrum [Fig. \ref{000_8}(a)]. A specific
feature of the strong coupling case occurs for relatively small
electric fields [see an enlarged picture plotted in Fig.
\ref{3strong}(b)] and consists in avoided crossings that occur
between energy levels of {\it different} families. In Fig.
\ref{3strong}(b) there are two such avoided crossings - marked by
dashed rectangles. In the ground state the avoided crossing involves
$1a, 2b$ and $3c$ bright energy levels. Electron and hole densities
for that avoided crossing are plotted in Figs. \ref{ff3strong}(m-r)
for $F=0$  and in Figs. \ref{ff3strong}(s-y) for $F=10$ kV/cm. At
$F=0$ all the states share nearly the same electron density [cf.
Fig. \ref{ff3strong}(m,o,q)] with a maximum in the central dot but
with a strong penetration to the other two dots.

\begin{figure}[htbp]
\hbox{\epsfysize=72mm
                \epsfbox[43 150 573 680] {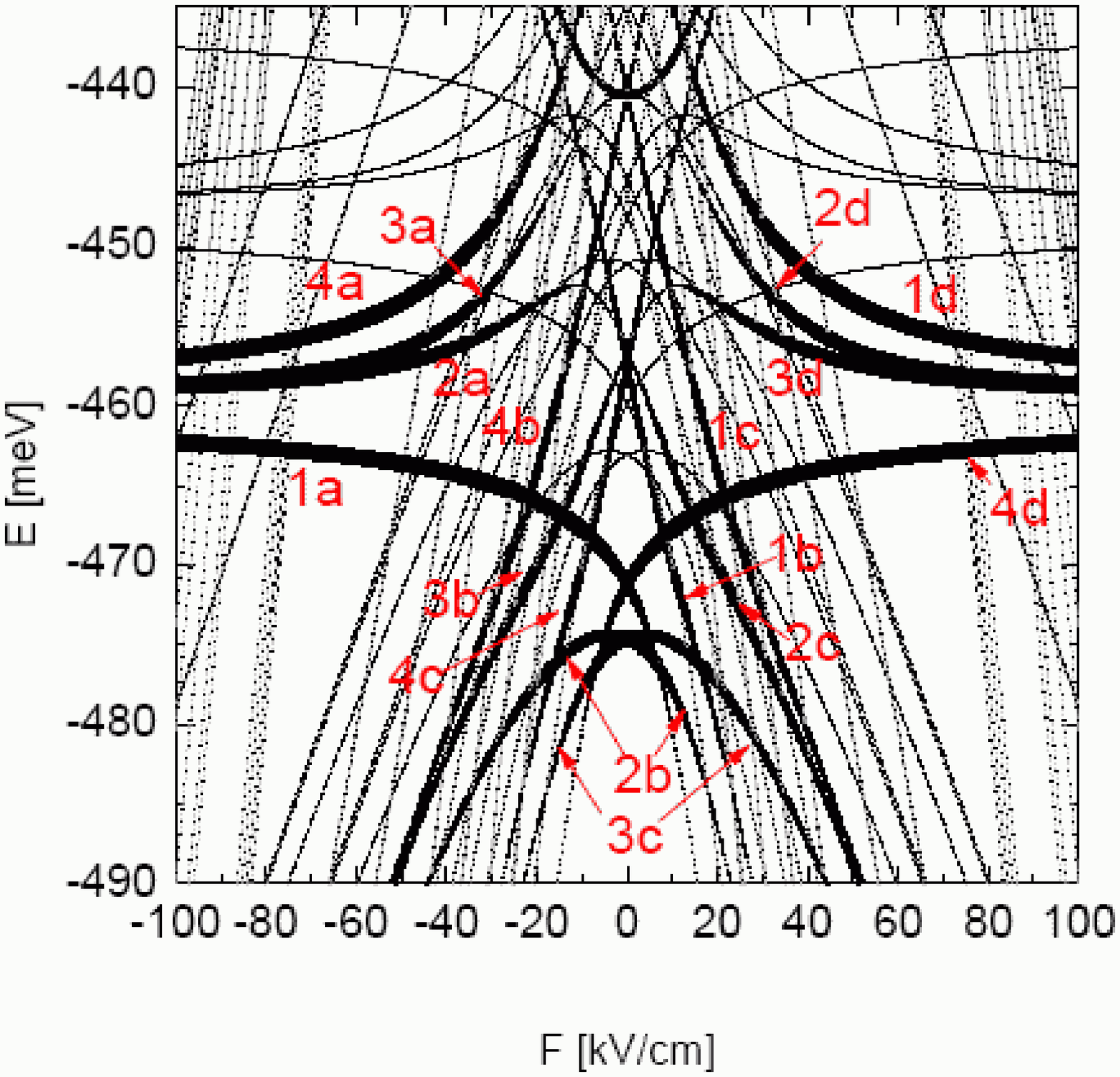}\hfill}
\caption{(color online) Same as Fig. \ref{000_8} (a) but for four
identical dots separated by identical barriers of thickness 4 nm.}
\label{0000_8}
\end{figure}

\begin{figure}[htbp]
\hbox{\epsfxsize=57mm
                \epsfbox[168 21 434 817] {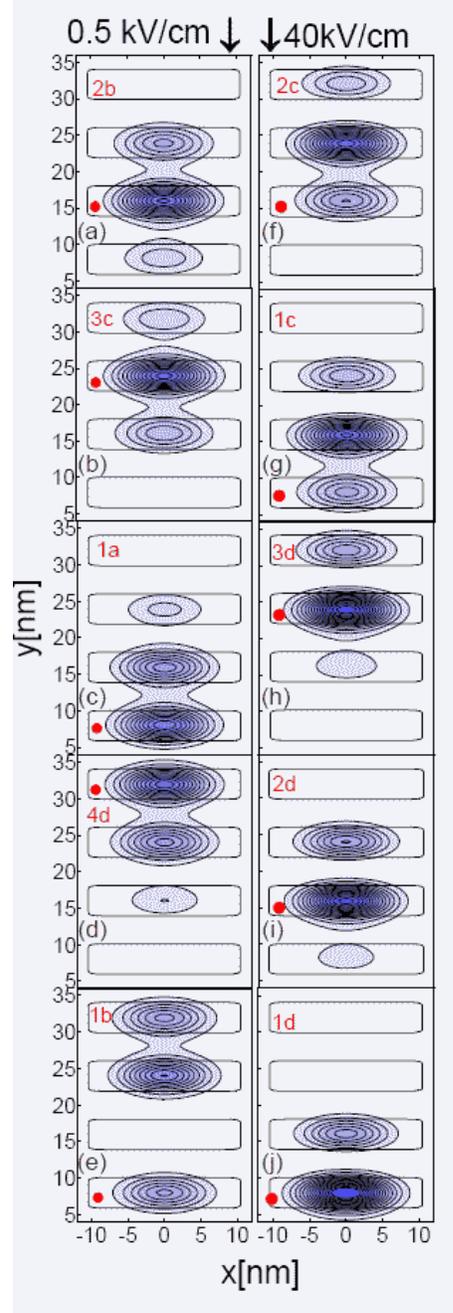}\hfill}
\caption{(color online) Electron densities for the stack of four
identical dots (for the energy levels presented in Fig.
\ref{0000_8}). Left column of plots (a-e) corresponds to $F=0.5$
kV/cm and the right column (f-j) to $F=40$ kV/cm. In each case the
hole is totally localized within the dot specified by the ``family''
label (additionally marked a red dot). }\label{0000_8ff}
\end{figure}

\begin{figure*}[ht!]\hspace{1cm}
\hbox{\epsfxsize=160mm
                \epsfbox[26 250 593 573]{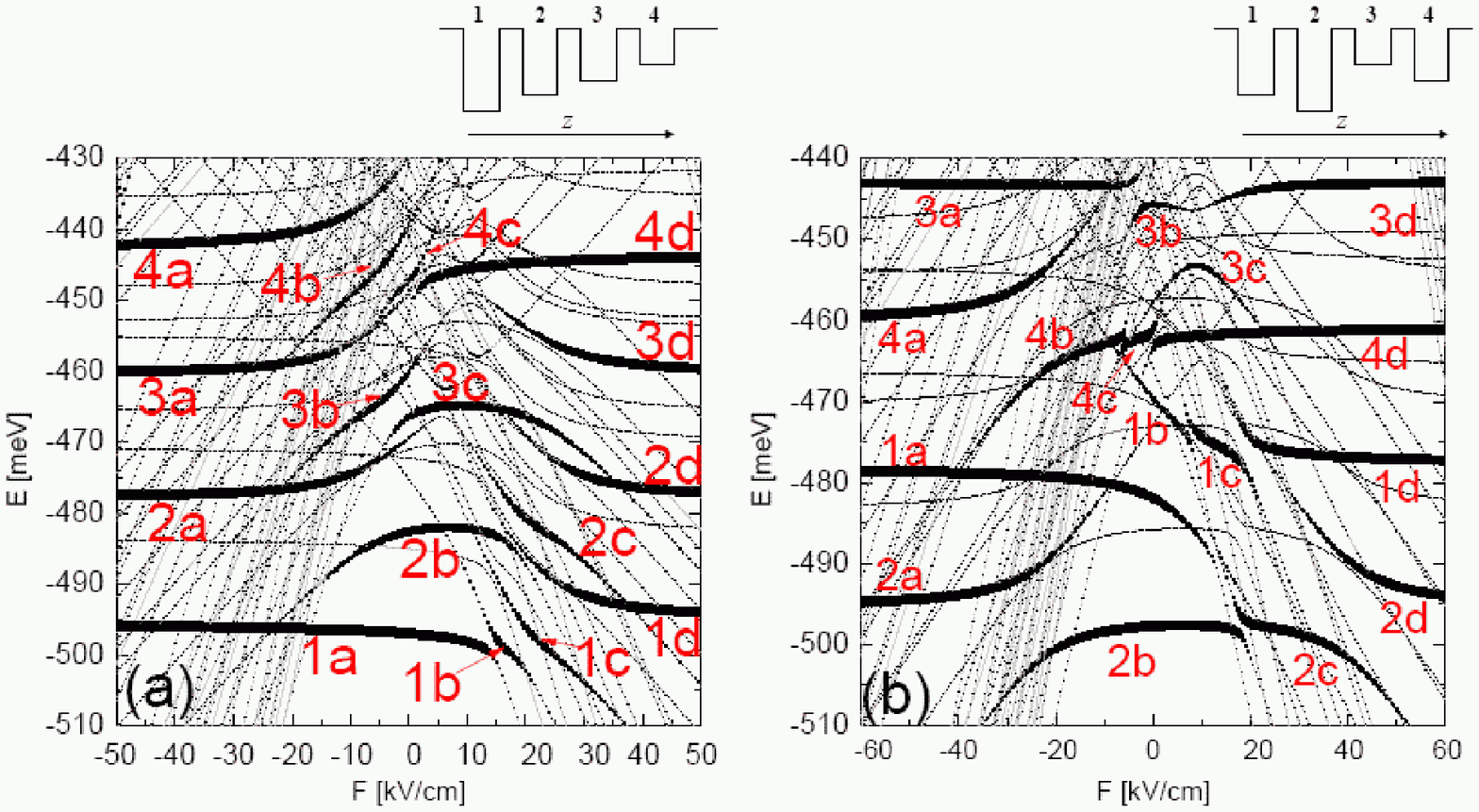}
                 \hfill}
\caption{(color online) The spectrum for four stacked dots for
barrier thickness of 5.5 nm. In (a) a constant 'gradient' of the
confinement potential depths of the dots is assumed: (see the
schematic drawing above the plot) \label{4k} the depth of each
subsequent dot is decreased by 10 meV for both the electron and the
hole when one moves up the stack. In (b) it is assumed that the
shifts of the dots depth for both carriers are from the bottom to
the top of the stack: +10, +20, -10 and 0 meV.}\label{4kpkiniei}
\end{figure*}

In the strong coupling limit a simple product of the electron and
hole single-particle wave functions [instead a linear combination of
such products - see Eq. (4)] is a relatively good approximation for
 the exciton wave function. Therefore, although in principle only
the total parity of the exciton state is a good quantum number at
$F=0$, in the strong coupling limit (and only in the strong coupling
limit) the states acquire also approximate single-particle parities
(as discussed in Ref. \cite{parity}). Below, we will refer to the
single-particle states and parities for the interpretation of the
avoided crossings related to the hole tunneling in the strong
coupling limit. In all the three states involved in the $1a-2b-3c$
avoided crossing [the lower rectangle in Fig. \ref{3strong}(b)] the
electron is in an even-parity state [see the electron density plot
in Figs. \ref{ff3strong}(m,o,q)], i.e. in its lowest-energy state.
In the ground-state the hole has a similar density [Fig.
\ref{ff3strong}(r)] but with a weaker penetration to the top and
bottom dots. The first excited state corresponds to a hole
excitation, i.e. to an antibinding hole state [Fig.
\ref{ff3strong}(p)]. At $F=0$ the hole in the first excited state
occupies an odd-parity energy level. Due to the odd parity the hole
density is bound to vanish in the center of the central dot. We see
that in fact the hole is totaly removed from the central dot [Fig.
\ref{ff3strong}(p)]. The total envelope exciton wave function for
this state is odd and therefore the recombination is strictly
forbidden since the integral (\ref{recprob}) vanishes. For a product
of single-particle wave functions $\phi({\bf r_e},{\bf r_h})
=f_e({\bf r_e})f_h({\bf r_h})$ the formula for the recombination
probability (\ref{recprob}) reduces to
\begin{equation} p=\left |\int d^3{\bf r}f_e({\bf
r}) f_h({\bf r})\right |^2 ,
\end{equation}
i.e., to the square of the overlap between the hole and electron
wave functions. For opposite parities of the two functions the
overlap is strictly zero. For the second excited energy level the
hole is an even state again [Fig. \ref{ff3strong}(n)] and the
recombination probability is therefore non-zero but turns out to be
small, about 15 times smaller as compared to the ground state. In
the second excited state at $F=0$ the hole wave function changes
sign in both barriers. The hole is more probable to be found at the
extreme dots of the stack while the electron wave function is
maximal in the central dot. This, along with the sign oscillations
of the hole wave function makes the electron and hole wave functions
nearly orthogonal, hence the small recombination probability. Note,
that the electron and hole densities presented in
 Fig. \ref{ff3strong}(m-r) are in a good qualitative agreement with
the calculations for $F=0$ presented in Fig. 5 of Ref.
\cite{Ledentsov}
 for the electron and hole states in a stack of three
strongly coupled dots each of different size and shape.

When the electric field is switched on the recombination probability
of the ground-state is reduced and all the three energy levels leave
the avoided crossing region [lower rectangle in Fig.
\ref{3strong}(b)] as bright states with comparable recombination
probabilities. The electron and hole densities for these energy
levels are plotted for $F=10$ kV/cm in Figs. \ref{ff3strong}(s-y).
Already for this field the hole tunnel coupling is nearly
extinguished [in Figs. \ref{ff3strong}(t,v,y)]. When the hole
becomes localized in the single dot, the discussion of energy levels
in previously used terms of families becomes valid again. Therefore,
outside the avoided crossing region we may identify the ground state
as the $1a$ energy level, the first excited bright state as $2b$,
and the second as $3c$. Outside the discussed avoided crossing range
of small electric fields, the energy levels of different families
cross like in the intermediate coupling case [see for instance the
crossing of the $3c$ and $1b$ levels in Fig. \ref{3strong}(b)].

Let us now turn our attention to the second avoided crossing
observed in Fig. \ref{3strong}(b) (the region marked with the upper
dashed rectangle). From the negative $F$ side the avoided crossing
occurs between the $1b$, $2a$, $3b$ energy levels. At $F=0$ in all
the states that enter this anticrossing the electron is in the first
excited state of odd parity with respect to the center of the stack
[see Figs. \ref{ff3strong}(a,c,e)]. For symmetry reasons, the
electron density in this state vanishes in the center of the stack,
but not entirely in the central dot, since a leakage from the
extreme dots is observed.  At $F=0$, in the lowest of the three
energy levels participating in the avoided crossing, the
recombination is forbidden due to the even parity of the hole
[opposite to the electron parity -- see Fig. \ref{ff3strong}(f)].
For the same reason, at $F=0$ the recombination from the
highest-energy of the three anticrossing levels is strictly
forbidden [for the hole density see Fig.\ref{ff3strong}(b)].
Consequently only the middle-energy state is bright at $F=0$. Note
the similarity of the electron and hole densities in the bright
state at $F=0$ [compare Figs. \ref{ff3strong}(c) and (d)] both the
particles are in the lowest-energy antibinding $l=0$ energy levels.
For $F=10$ kV/cm the hole coupling is removed [see Figs.
\ref{ff3strong}(h,j,l)]. The order of the bright energy levels when
they leave the avoided crossing region is the same as in the
ground-state avoided crossing, i.e. the lowest-energy bright level
at $F>0$ is associated to the hole localized in dot $1$, the middle
energy - with dot $2$ and the energy level in which the hole is in
the dot $3$ goes up with growing $F$.

In the intermediate coupling region, when we varied the dot
confinement potential, the families of energy levels shifted with
respect to each other, as discussed in subsection III.B. The exciton
spectrum for the case in which the potential depths are changed
according to the ``constant gradient'' case (as discussed in III.B)
is plotted in Fig. \ref{3strong}(c). Comparison of Figs.
\ref{3strong}(a) and \ref{3strong}(c) leads to the conclusion that
the effect of the confinement variation on the spectrum for the
strong coupling is distinctly less pronounced than for the
intermediate coupling case. This is due to the electron coupling
effects that are much stronger as compared to the potential
variation (see the begining of Section III). In this sense the
differences in dot confinement are masked by the strong electron
tunnel coupling. The avoided crossings that appear for small fields
due to the hole tunnel coupling are presented in detail in Fig.
\ref{3strong}(d). For nonidentical dots they are no longer symmetric
and they are shifted with respect to $F=0$ field, but they
distinctly preserve their character [compare with the identical dots
case of Fig. \ref{3strong}(b)]. In particular, in both avoided
crossings marked by rectangles in Fig. \ref{3strong}(d) at the
center of the avoided crossing a single energy level with a much
stronger recombination probability than the others appears. Also the
energy position of the brighter energy level within the triple of
interacting energy levels is conserved. In the lowest-energy avoided
crossing it is the lowest energy level. In the second avoided
crossing the brightest is the second energy level of the three
interacting levels. Recombination of the two other levels is no
longer forbidden by the symmetry-related selection rules but it
remains small. Same rule applies for the next avoided crossing on an
energy scale that did not fit into Fig. \ref{3strong}.

\begin{figure}[ht!]
\hbox{\epsfxsize=80mm
                \epsfbox[143 10 425 820] {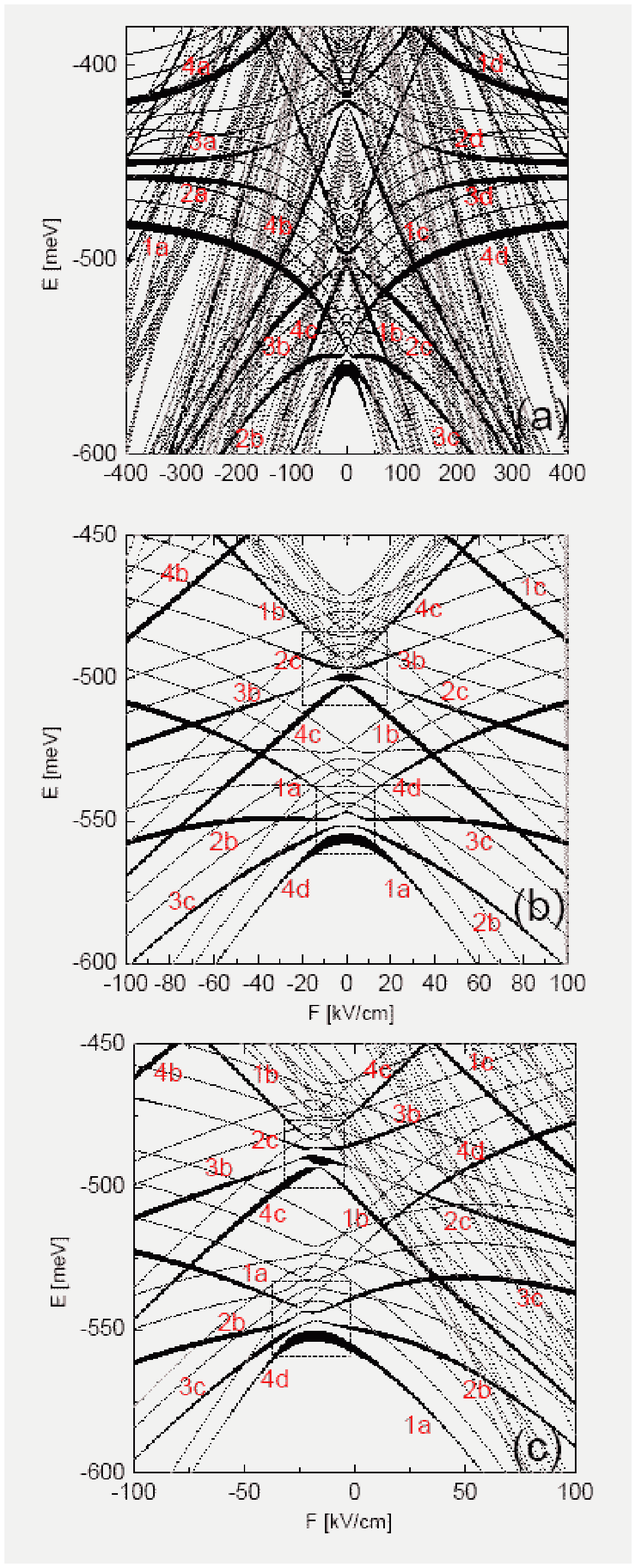}
                \epsfysize=70mm\hfill}
 \caption{(color online) Spectra for four dots with
barrier width of 1 nm; (a,b) correspond to identical dots and in (c)
the depth of the confinement potentials of the dots 1--4 is varied
by +10, +0, -10 meV, and -20 meV, respectively. (b) is a zoom of
(a). } \label{4strong}
\end{figure}

\begin{figure}[ht!]
\hbox{\epsfysize=162mm
                \epsfbox[74 28 493 826] {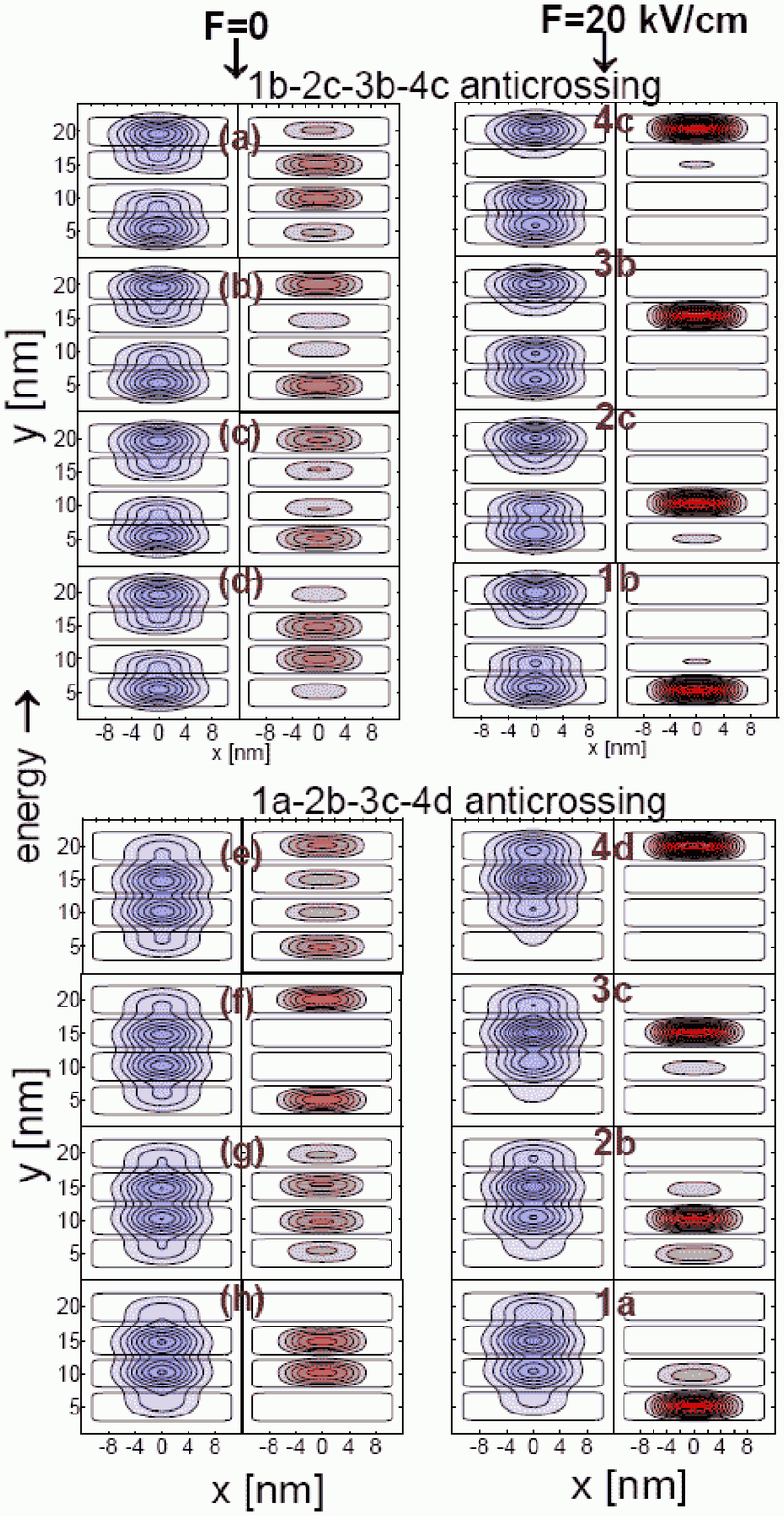}\hfill}
\caption{(color online) Cross section ($y=0$) of the electron and
hole densities corresponding to the energy levels participating in
the two avoided crossings marked by rectangles in Fig.
\ref{4strong}(b). Lower group of plots [the ones marked by (e-h) for
$F=0$, and by the family labels $1a$, $2b$, $3c$, $4c$ for $F=20$
kV/cm] corresponds to the ground-state avoided crossing marked by
the lower rectangle in Fig. \ref{4strong}(b). The upper group of
plots [(a-d) and $1b$, $2c$, $3b$ and $4c$]) corresponds to the
avoided crossing marked by the upper rectangle in Fig.
\ref{4strong}(b). In each group - the higher energy states
correspond to the upper panels. The first (second) column shows the
electron (hole) density at $F=0$ and the third (fourth) the electron
(hole) density at $F=20$ kV/cm. Identities of the states when their
energy levels emerge from the avoided crossing can be deduced from
the plots for $F=20$ kV/cm. The corresponding densities are
therefore labeled in the ``family'' notation.} \label{ff4strong}
\end{figure}

\subsection{Four identical dots -- intermediate coupling}
The spectrum for four identical dots separated by barriers of width
4 nm is presented in Fig. \ref{0000_8}. At $F=0$ the ground state is
twofold degenerate (like in the case of two-dot stacks -- see Fig.
\ref{2e} -- and in contrast to the stack of three dots -- see Fig.
\ref{000_8}). At $F=0$ in the ground state a crossing of $2b$ and
$3c$ energy levels is obtained as a function of electric field. This
degeneracy is due to the absence of hole tunneling and from the fact
that for the electron both internal dots are equivalent. The
electron densities in $2b$ and $3c$ energy levels at a weak electric
field of $F=0.5$ kV/cm can be inspected in Fig. \ref{0000_8ff}(a-b),
in which the dot occupied by the hole is marked by a red dot. In
both the $2b$ and $3c$ states the electron is most probable to be
found in the same dot as the hole. Electron tunneling to adjacent
dots is also observed but with a clear preference for the neighbor
dot that is situated inside the stack and not at its end. At a
positive electric field the $2b$ level goes down in energy (cf. Fig.
\ref{0000_8}) since the dipole moment of this state is aligned with
the electric field [see Fig. \ref{0000_8ff}(a) -- the centers of the
electron and hole distributions are displaced with respect to each
other in the direction of the electric force]. Opposite is the
effect of the field on the $3c$ level in which the dipole moment is
inverted. Due to the carriers distribution presented in Figs.
\ref{0000_8ff}(a,b) both these states have nonzero intrinsic dipole
moments, hence the shifts of the maxima of their energy levels off
$F=0$: to negative $F$ for $2b$, and to positive $F$ for $3c$.

 At $F=0$ the first excited state
is two-fold degenerate ($1a$ and $4d$ energy levels - see Fig.
\ref{0000_8}) and corresponds to the hole entirely localized in the
extreme dots of the stack. The electron density in these states is
maximal in the extreme dots but
 extends to the interior of the stack [see Fig. \ref{0000_8ff}(c,d)].

The electron densities plotted in Figs. \ref{0000_8ff}(f-j) for
$F=40$ kV/cm correspond to the energy levels $3d$, $2d$, $1d$ which
remain bright at $F>>0$ as well as  $1c$ and $2c$ levels which
appear as temporarily bright within the avoided crossings of
$1b$-$1d$ and the $2b$-$2d$ energy levels, respectively.

Comparing the four-dot spectrum of Fig. \ref{0000_8} with its
counterpart for three dots (Fig. \ref{000_8}) we notice that in the
$1x$ family -- associated to the dot at the end of the stack --
addition of a fourth dot results in the appearance of a second level
($1c$ in addition to $1b$) that depends nearly linearly on the field
and becomes bright when entering the avoided crossing between the
energy levels that remain bright at strong negative and positive
fields (in the four dot case these are labeled by $1a$ and $1d$,
respectively).

 For
three dots a member of the $2x$ family ($2b$) associated with the
internal dot of the stack was the non-degenerate ground state at
$F=0$. It entered a single avoided crossing with $2a$ [the bright
level at $F<<0$] and $2c$ [the bright one at $F>>0$]. Consequently
$2b$ level was a symmetric  function of $F$. For four dots in the
$2x$ family an energy level appears that is temporarily bright at
$F>0$ with a nearly linear dependence on $F$. For four dots this
level is labeled by $2c$. Due to its interaction with the $2c$ level
the 'parabolic' $2b$ energy level looses its symmetry with respect
to $F=0$ and becomes a deformed parabola as a function of field.

\subsection{Four non-identical dots -- intermediate coupling}

Fig. \ref{4k}(a) shows the spectrum for a stack of four dots with a
``gradient'' of the depth of the confinement potential  -- the
depths of the dots from the lowest to the uppermost are increased
for both carriers by 20, 10, 0 and $-10$ meV, respectively  [see the
schematic drawing of the confinement potential above Fig.
\ref{4k}(a)]. Near $F=0$ the ground state is the bright energy level
with the carriers confined in the deepest dot (number 1). The
ground-state dissociation appears with $1a$-$1b$ avoided crossings
near $F=15$ kV/cm.  For higher electric field the ground state is
dark and corresponds to the electron in dot 4. The lowest bright
energy level between 15 and 20 kV/cm is $1b$. For $F>20$ kV/cm, the
electron in $1b$ becomes localized in dot 3 and the energy level
becomes dark. We notice the reduced slope of the energy level $1b$
when it becomes dark compared to the dark ground state. The
subsequent bright energy level $1c$ becomes dark with the electron
localized in dot 2, that is adjacent to the dot occupied by the
hole. Consequently the dipole moment (slope of the energy level as
function of $F$) is still smaller. For the same reason we notice
that in $1c$ the electron-hole pair is significantly more resistant
to dissociation as compared to $1b$ (note the increased $F$ interval
in which $1c$ is bright).

In the $1x$ family all the avoided crossings of the bright energy
levels appear at positive $F$, which tends to remove the electron
from the lowest dot. On the other hand in the $2x$ family the
electron is removed by negative $F$ to dot 1 producing a wide $2a$ -
$2b$ avoided crossings (similarly as in the stack of three dots)
near $F=-10$ kV/cm. For positive field two avoided crossings are
observed. The first one $2b$ - $2c$ appears when the electron is
removed to dot 4, and the second one $2c$ - $2d$ when the electron
is removed to dot 3. Note that the slope of the dark energy level
that comes from the $2b$ - $2c$ avoided crossing is parallel to the
dark part of the $1b$ energy level. In both cases the hole and
electron are separated by an empty dot. Similarly,  the $2c$ energy
level when dark is parallel to the dark $1c$ energy level, since the
carriers occupy adjacent dots.

The avoided crossing patterns in the $4x$ family is clearly an
inverted and shifted pattern of the $1x$ family. The inversion is
due to the hole localized at the other extremity of the stack and
the shift to the reduced depth of dot 4. Similarly, the $3x$ family
is a shifted and inverted counterpart of the $2x$ family.

Fig. \ref{4k}(b) shows the spectrum for varied depths of the dots.
Dot 2 is deepest with depth increased by 20 meV for both carriers
from the values given in Section II (see the schematic potential
profile above the figure).  The analogous shifts for dots 1, 3 and 4
are 10, -10, and 0 meV, respectively.  The patterns of the avoided
crossings remain characteristic to the position of the dot in which
the hole is localized within the stack [cf. Fig. \ref{4k}(a)]. More
pronounced deformation appears only in the $3x$ family associated
with the dot which is the shallowest and is visible in a modified
shape of the $3b$ bright line. The neat correspondence (shift and
inversion) between families $1x-4x$ and $2x-3x$ from the constant
gradient case of Fig. \ref{4k}(a) is not conserved, but the families
of energy levels can still be recognized in the spectrum.

\subsection{Four dots -- strong coupling}
Similarly to the three dot case, the spectrum for four dots in the
strong coupling regime [see Fig. \ref{4strong}(a)] resembles the one
of the intermediate coupling case (see Fig. \ref{0000_8}). Like for
three strongly coupled dots (subsection III.C) qualitative
differences appearing for small $F$, where avoided crossings of
states of different families are obtained. The region of the
two-lowest avoided crossing is shown in more detail in Fig.
\ref{4strong}(b) with the hole and electron densities plotted in
Fig. \ref{ff4strong}. In the ground state we have avoided crossing
of $1a$, $2a$, $3c$ and $4d$ energy levels. In the center of the
avoided crossing (here  $F=0$) the ground-state becomes distinctly
brighter than the three other levels. The hole and electron density
in these states look alike [see Fig. \ref{ff4strong}(h)], with the
exception of weaker hole tunneling. As a matter of fact at $F=0$ the
electron densities are nearly identical for all the four energy
levels [see Fig. \ref{ff4strong}(e-h)]. In all these energy levels
the electron is in an even-parity single-particle state. In the
first excited state the hole state is odd, so recombination is
forbidden [see Fig. \ref{ff4strong}(g) and Eq. (8)]. For the same
reason it is forbidden for the third excited state [see Fig.
\ref{ff4strong}(e)]. The recombination is allowed for the second
excited state [see Fig. \ref{ff4strong}(f)], in which the hole is in
an even parity state. The recombination probability is nonzero but
small (12 times smaller than in the ground state) since the hole
occupies only dots at the top and the bottom of the stack, and the
electron occupies mostly the internal ones.

An electric field of $F=20$ kV/cm is nearly strong enough to break
the hole tunnel coupling, which results in the hole localization in
a single dot.

In the second avoided crossing that appears higher in energy and
involves the $1b$, $2c$, $3b$ and $4c$ energy levels, the electron
at $F=0$ occupies an odd parity state with probability maxima at the
ends of the stack [see Figs. \ref{ff4strong}(a-d)]. The parity of
the hole in the subsequent energy levels change like in the lower
avoided crossing: it is even in the lowest energy state, odd in the
second-energy state, etc. Recombination is not forbidden by symmetry
from the second- [Fig. \ref{ff4strong}(b)] and fourth- [Fig.
\ref{ff4strong}(a)] -energy states. It is the second energy level
which is the brightest, since then the hole density is maximal in
the outer dots like the electron density. In this state both the
electron and the hole are in their first single-particle excitated
states. As a general rule the largest recombination probability is
obtained for the cases that the electron and the hole excitation are
the same. As an additional illustration, in Fig. \ref{4strong}(a) we
notice that in the third avoided crossing (associated with the third
single-electron state) it is the third-energy state that is the
brightest at $F=0$.

The low-field spectrum for a system of nonidentical dots is
presented in Fig. \ref{4strong}(c). The lowest dot is increased by
20 meV, the second by 10 meV and the uppermost is decreased by 10
meV (``constant gradient''). The avoided crossings related to mixing
of the families are shifted from $F=0$ and deformed, i.e. not
symmetric with respect to the center of the avoided crossing. For
instance in the lowest-energy bright state the left arm tends to the
$4d$ energy level when the hole tunneling is lifted and the right
arm tends  to $1a$. For identical dots $4d$ and $1a$ are equivalent
counterparts -- the energy of $4d$ at $F$ is equal to the energy of
$1a$ at $-F$. This is no longer the case for non-identical dots,
hence the asymmetry in the avoided crossing. The asymmetry is also
observed in the recombination probability. In particular the
recombination probability of energy level $1a$ vanishes with field
in a slower way than in the $4d$ energy level, which is a direct
consequence of the difference in the depths of dots $1$ and $4$. In
spite of the asymmetry in the avoided crossings the general features
are similar to those observed for identical dots. A single energy
level much brighter than the others appears in the center of the
avoided crossings.

\section{Discussion}

Our results for intermediate coupling of $N$ identical dots show
that the exciton spectrum can be divided into $N$ families, each
associated with the localization of the hole in one of the dots
within the stack. The energy levels of different families cross
while avoided crossings are observed between bright levels of the
same family as function of the external electric field which
transfers the electron within the stack of $N$ dots. Each family
contains $N$ energy levels that appear as bright for a certain
electric field range. In each family we have a single energy level
which remains bright even at strong negative electric field $F<<0$
and another that stays bright for $F>>0$. Below we refer to these
energy levels as ``ultimate''. These two energy levels undergo an
avoided crossing around $F$ close to zero. Below we call this
avoided crossing as the ``principle'' one. For previously studied
case of an exciton in two
dots\cite{ek1,ek2,ek3,ek4,xk1,xk2,doty,small} there are no other
bright energy levels than the ultimate ones. However, for $N\ge 3$,
the remaining $N-2$ levels of the family become bright in a finite
electric field range contained within the principle avoided
crossing. These temporarily bright energy levels depend on the
external field in a much stronger way than the ultimate bright
levels in the strong field limit. In the limit of a strong field,
when both the electron and the hole are localized in the same dot,
the $F$-dependence of the ultimate levels is only due to a
single-dot exciton polarizability.\cite{barker} On the other hand,
the intrinsic and induced dipole moments of the temporarily bright
energy levels are much larger, since they appear in the spectrum
when the electron is removed from the dot in which the hole is
localized.  For the families of energy levels associated to the dots
at the extreme ends of the stack all the $N-2$ temporarily bright
levels depend on the electric field nearly linearly. Their slope is
related to the dipole moment of the electron-hole pair when it
becomes dissociated [this is most clearly visible in Fig.
\ref{4kpkiniei}(a)]. For the families associated with the dot
situated inside the stack a {\it single} temporarily bright energy
level strongly deviates from the nearly linear dependence on $F$.
Its dependence on the field resembles a parabola (deformed for odd
$N$) with arms pointing down on both sides of its maximum located
close to $F=0$. Position of the dot inside the stack with which the
'parabolic' level is associated can be deduced by counting the
number of the temporarily bright energy levels that are nearly
linear in $F$ and enter between this 'parabolic' energy level and
the ultimate bright energy levels at $F<<0$ and $F>>0$. Namely, the
number of bright linear energy levels appearing at the positive
field side of the maximum of the 'parabolic' bright energy level is
equal to the number of dots between the hole-containing-dot
(defining the family) and the extreme dot at the top of the stack,
and conversely for $F<0$. Thus, addition of a $N$-th dot to a stack
containing $N-1$ dots results in the appearance of a {\it single}
new temporarily bright level in each of the pre-existing families.
This new level depends nearly linearly on the electric field.
Obviously, a new family is formed with the added dot. When a dot is
added to the stack the number of internal dots is increased by one
and a single new temporarily bright 'parabolic' (and not linear)
energy level appears in the spectrum of the entire stack. For
instance, when the third dot is added to a stack of two dots the
first 'parabolic' level appears -- the one labeled by $2b$ in Fig.
\ref{000_8}. For $N=4$ we have already two such parabolic levels:
$2b$ and $3c$ -- see Fig. \ref{0000_8} -- which cross at $F=0$.

For the intermediate coupling the energy levels of the hole alone
are $N-$ fold degenerate. This degeneracy is lifted by the
interaction with the electron which tends to be localized in the
center of the stack, where we have a single dot for odd $N$ and a
couple of dots for even $N$. This provides a general rule that at
$F=0$ for $N$ identical dots in the intermediate coupling the
ground-state is non-degenerate for odd $N$ and two-fold degenerate
for even $N$.

For strongly coupled dots hole tunnel coupling becomes active.
According to our results, even in the extreme case of 1 nm - thin
barrier the coupling is removed by a relatively weak electric field
of about 20 kV/cm. When the hole coupling is removed by the electric
field, the discussion of the spectrum in terms of families of energy
levels associated to the hole localized in one of the dots is valid
again. However, mixing of the hole states is found for small values
of the field in the avoided crossing in which $N$ bright energy
levels participate with a single member of each family. The
participating energy levels are associated to the same
single-particle state of the electron. Due to the strong electron
tunnel coupling, within the range of each of these avoided
crossings, the electron localization is only weakly perturbed by the
electric field, and this is the hole which is redistributed between
the dots by the field. This is the inverse mechanism to the one
observed for the avoided crossings in the intermediate coupling
regime, where the localization of the hole is fixed and only the
electron localization within the stack is changed by the electric
field. The avoided crossings related to the hole tunneling and
mixing between the families have two characteristic features: 1) In
the center of each avoided crossing a single energy level becomes
much brighter at the expense of the recombination probabilities of
the other participating levels. A distinctly larger recombination
probability is obtained for the hole state which is compatible with
the state of the electron, i.e., corresponds to the same excitation
in the direction parallel to the axis of the stack. This results in
the rule, that {\it within} the avoided crossing that is $m-$th in
the energy scale (i.e. associated with the $m-$th excitation of the
electron) it is the $m-$th of the participating energy levels which
becomes the brightest in the center of the avoided crossings. 2) The
order of the bright energy levels as they reappear outside the
avoided crossing depends only on the position of the specific dot
within the stack in which the hole becomes trapped. On the positive
$F$ side of the avoided crossing, lower energies corresponds to the
hole localization in the lower dots [$F>0$ pushes the hole down --
see Fig. \ref{schema}(a)]. The avoided crossings that are due to the
hole transfer -- with the spectacular modulation of the
recombination probabilities -- are only deformed by the confinement
variation. The order of the subsequent avoided crossings in energy
remains unchanged by the dot variation since their character is
defined by the state of the electron and the electron energy
splitting is huge when the dots are close enough for the hole to
tunnel.

In the intermediate coupling regime the variation of the depths of
the dot within the stack leads to energy shifts of the families of
energy levels. The families are also translated on the electric
field scale. Nevertheless, the pattern of the avoided crossing
remains characteristic to the position of the hole-containing dot
within the stack. In the strong coupling regime the variation of the
confinement depths within the stack of the order of 10 meV is much
smaller than the coupling-related splitting of electron energy
levels. Then, the difference of the confinement depths may not be
resolved by the electron so the spectra of the strongly coupled dots
are less sensitive to the confinement variation along the stack [cf.
Fig. \ref{3strong} for strongly coupled triple of identical dots (a)
or non-identical dots (c)].

\section{Summary and conclusions}
We calculated the exciton spectra in triple and quadruple vertically
stacked self-assembled quantum dots in the presence of an external
electric field using the configuration interaction approach.
Intermediate and strong coupling regimes were considered and we
explained how the spectra evolve when an additional dot is added to
the stack.

In the intermediate coupling regime the bright energy levels can be
separated into families each associated with a specific dot of the
stack in which the hole is localized. The electron transfer between
the dots induced by the electric field is associated with the
appearance of a pattern of avoided crossings that is characteristic
for each family of energy levels. The structure of the avoided
crossing is related mainly to the position of the dot within the
stack which is translated into the number of avoided crossings
observed for opposite electric field orientations. We found that the
pattern of avoided crossing is related to the electron transfer
between the dots which remains qualitatively similar when the depths
of the dots are varied. The depth variation results in relative
shifts of the families of energy levels. Therefore,
photoluminescence experiment should be able to probe the variation
of the depth of the effective confinement potential along the stack.
In particular, the experimental observation of the pattern of
avoided crossings for the lowest bright energy levels should
indicate which of dots is the deepest within the stack.

We discussed the avoided crossings between levels of different
families that appear for strongly coupled dots. These anticrossings
appear at small $F$ and involve $N$ bright energy levels, where $N$
is the number of the dots within the stack. Within each anticrossing
one of the energy levels increases in brightness at the expense of
$N-1$ others. In the brightest state the hole occupies a
single-particle orbital which is compatible with the orbital of the
electron, i.e. corresponds to the same single-particle state. The
mixing between families occurs only for relatively weak electric
fields. For stronger fields the hole is localized in a single dot
and the pattern of avoided crossings becomes qualitatively similar
to the intermediate coupling case.

 \acknowledgments This work
was supported by the EU Network of Excellence: SANDiE and the
Belgian Science Policy (IAP).


\begin{thebibliography}{00}
\bibitem{odnosnik} A small electric field of
1kV/cm was assumed in Fig. \ref{ff000_8}(a,b) to lift the parity
symmetry of the states. In a system of symmetric dots, due to the
parity symmetry, at $F=0$ the hole resides equally in all the dots
in spite of the absence of interdot hole tunelling.
\end{thebibliography}
\end{document}